\documentclass[12pt]{iopart}
\usepackage{iopams} 
\usepackage{bm}
\usepackage{graphicx}
\begin{document}

\title[]{Semiclassical triton}

\author{Nishchal R. Dwivedi$^{1,2}$, Harjeet Kaur$^{3*}$ and Sudhir R. Jain$^{2,4}$}

\address{University of Mumbai, Mumbai-400 098, India$^{1}$,\\
Nuclear Physics Division, Bhabha Atomic Research Centre, Mumbai-400 085, India$^{2}$,\\
Department of Physics, Guru Nanak Dev University, Amritsar-143005, India$^{3}$,\\
Homi Bhabha National Institute, Anushakti Nagar, Mumbai-400 094, India$^{4}$.}
\ead{harjeet\_kaur17@yahoo.com}
\begin{abstract}
 The symmetric components of the spatial part of $S$- and $D$- states' wavefunctions for triton $(^{3}H)$ are investigated utilizing semiclassical expansion (in the powers of $\hbar$). Analysis of the diagonalized Hamiltonian reveals the existence of two different mass states within the ground state of triton. We have solved the coupled differential equations for the two admixed states $^{2}S_{1/2}$ and $^{4}D_{1/2}$ owing to tensor interactions  exploiting classical WKB-theory using phenomenological Feshbach-Pease potentials. The relative probability of the $D$-state is found to be in good agreement with the experimentally inferred value (4 - 5 \%).

\end{abstract}


\maketitle

\section{\label{sec:level1}Introduction}

The presence of non-central force among nucleons complicates the nature of bound states for even the simplest nuclei \cite{blatt}. This manifests in, for instance, deuteron which is deformed in its ground state. The admixture of ${}^3 S_{1}$ and ${}^3 D_{1}$ states in deuteron can be exactly understood within a semiclassical formalism \cite{Jain2004} based on the coupled equations set up by Rarita and Schwinger \cite{blatt}. As soon as we take the next step, and try to embark on an understanding of the nucleus with next mass number, we encounter a three-body problem. Here, we deal with triton where a proton binds the two neutrons along to yield a ground state with spin and parity, $J^{\pi} = \frac{1}{2}^+$ and isospin $T=\frac{1}{2}$. Due to the fact there are three particles, we need to carefully specify their positions and the arrangements (permutations). Whereas in the case of deuteron, the tensor interaction ${\bf{S_{12}}}$ mixes only $S$- and $D$- components, triton can support a much more elaborate set of components \cite{blatt}. Symmetric $S$-state is the dominant component in the ground state while $D$- and $P$- states have lesser contributions. We follow Derrick and Blatt \cite{Derrick1958} in order to simplify the analysis of the triton problem. The main achievement of this work was to reduce the analysis to a set of coupled equations. We present the solutions of these equations in a general manner, following \cite{Jain2004}. The method involves repeated usage of unitary transformations, where the results become better by an order of $\hbar$ after each application. 

As there has been a lot of significant work done on the triton wavefunctions over the years, it is imperative that the present work is placed in perspective. Exact solutions for the bound state of triton was found in the pioneering work by Mitra \cite{mitra}  who solved the Schr\"{o}dinger equation using separable potential. Although Jacobi coordinates were argued to be inappropriate for the nuclear three-body problem \cite{Derrick1958}, they were employed to obtain radial wavefunction \cite{gibson67}  which could explain trinucleon elastic scattering form factors and two-body photodisintegration cross-sections. Numerical solution of the configuration-space Faddeev equations yielded wavefunction probabilities for s-wave interaction model and the tensor force model of Reid \cite{gibson85}. An updated review of work until the early 1990s can be found in \cite{wu}.  All these works and the ab initio methods are indeed powerful and they have been useful in understanding many aspects of the nuclear three-body problem. However, they are all numerical. 

The discussion we present here is analytic for the most part, until we get down to finding an integral or plotting a function, to put it rather simply. More than ten years ago, the deuteron problem was solved in a similar manner, leading to two mass states corresponding to two possible degenerate combinations of $S$- and $D$- states - a predominantly $S$- state with small fraction of $D$- component, and, a predominantly $D$- state with a small $S$-component - obviously, as the $D$- state is weaker, the second combination is much weaker than the first one. This is in line with the discussion attributed to Schwinger in \cite{blatt} (page 111 where they are called $\alpha$ and $\beta$ states). However, what could not be seen until a systematic analysis was presented in \cite{Jain2004} that these two states are mass states. Thus, our present work is a continuation of the work on deuteron on what we call as semiclassical nuclear dynamics, now for the three-body problem. It turns out that the method works here also very well. The solvability of this problem by a method where the results are expressed as an expansion in powers of Planck's constant is particularly exciting for us to explore connections with quantum chaos.

Three-body problems are also of a great interest as they generically exhibit classically non-integrable behaviour \cite{berry}. The presence of nonlinear resonances due to the perturbation of a two-body problem shows in a remarkable manner in the patterns of rings of Saturn or belt of asteroids between Mars and Jupiter, as also in the motion of the Moon  \cite{gutzwiller}. In the case of the nuclear three-body problem, the effects are non-perturbative. Thus we resort to a semiclassical method for matrix Hamiltonians \cite{littlejohn, gaspard}. Semiclassical methods have the potential of providing exact solutions to the long standing problems \cite{brack2003,Jain1998,Jain1995}. By applying trace formulae, melting of shell effects in nuclei at high excitations has been understood recently \cite{Kaur}. The nature of nucleonic motion is quite intricate and fascinating. It will be seen that the setting in which we solve this problem brings out that there is a three-nucleon motion in the complex plane where the bound state is built as the system oscillates and tunnels across potential energy surfaces. The mixing of important components consisting the ground state wavefunction has been incorporated in an exact manner.

\section{Structure of the wavefunction for triton}

There are nine coordinates specifying the positions of the particles. Fixing the centre of mass (CM) reduces them to six coordinates. We can imagine that these six coordinates specify the three sides of a triangle denoted by $\bf{r_{12}}$, $\bf{r_{23}}$ and $\bf{r_{31}}$ and three Euler angles relative to a standard orientation (the standard orientation has the CM at the origin and one particle positioned on the $x$-axis in the plane-$XY$ containing the triangle). Following \cite{Derrick1958}, the group classification helps simplifying the wave equation by eliminating the Euler angles.

The wave equations can be written as a product of three factors - originating from the sides of the triangle, the Euler angles, and spin-isospin. The factors depending on the Euler angles, spin and isospin can be combined by using the Clebsch-Gordan coefficients. These combinations, owing to the representations of the symmetric group for three elements, could have a symmetric, anti-symmetric or a mixed form. 

The total antisymmetric wavefunction for $S-$ and $D-$ states of the triton is given as:
\begin{eqnarray}
\psi &=& f_1 \mathcal{Y}_{1}+ f_{13}\left[\frac{1}{\sqrt{3}}\left(F\mathcal{Y}_{8,2}+ G \mathcal{Y}_{8,1}\right)+ \mathrm{cosec} \lambda \left(- F  \mathcal{Y}_{9,2}- G   \mathcal{Y}_{9,1}\right)\right. \nonumber\\
&+& \cot \lambda  \left(- F  \mathcal{Y}_{10, 2}- G  \mathcal{Y}_{10,1}\right)\bigg].
\label{wave}
\end{eqnarray}

The parameters ($F,G,\lambda$) in terms of the interparticle distances are given as follows:
\begin{eqnarray}
F&=&\frac{r_{23}^2+r_{13}^2-2r_{12}^2}{\sqrt{3}},~G=r_{23}^2-r_{13}^2,~\cos \lambda =\frac{4\sqrt{3}\Delta}{R^2},\nonumber\\ 
~\Delta &=&\frac{1}{2}|r_{31}\times r_{12}|, ~ 
 R^2=r_{12}^2+r_{23}^2+r_{13}^2,~ 0\leq \lambda \leq \pi/2. 
\end{eqnarray}
Only symmetric component of the spatial wave functions $f_{1}$ and $f_{13}$ are chosen because for the antisymmetric configurations, under the exchange of space-coordinates of particles $1$ and $2$ we would have $\psi = 0$ whenever $r_1 = r_2$. Such wavefunctions vary rapidly in space and do not form energetically favourable states. 

However, symmetric components are normalized as follows:
\begin{eqnarray}
\int  d\tau \left[f_{1}^2+N_{5}f_{13}^2\right]=1, ~ \mathrm{where} ~ 
N_{5} =R^{4}\left(\frac{2}{3}-\frac{2}{9}\sin ^2 \lambda \right)~ \mathrm{and}~ \nonumber\\
 \int  d \tau = \int_{0}^{\infty}dr_{23}\int_{0}^{\infty}dr_{13}\int_{|r_{23}-r_{13}|}^{r_{23}+r_{13}}dr_{12}r_{23}r_{13}r_{12}. 
 \end{eqnarray}
$\mathcal{Y}_{1},\mathcal{Y}_{8,1},\mathcal{Y}_{8,2},\mathcal{Y}_{9,1},\mathcal{Y}_{9,2},\mathcal{Y}_{10,1}$ and $\mathcal{Y}_{10,2}$ are the total angular momentum-isospin wave functions obtained by the linear combinations of products of the functions of Euler angles, spin and isospin  which transform under ``complete" rotations and joint Euler angle-spin-isospin coordinate permutations according to symmetric group $S_3$. Euler angle wavefunctions $Y_{L}^{M_{L}}(P_{e},\mu)$ are characterized by the orbital angular momentum $L$, $\mu$ (the body $z$-component of the orbital angular momentum), $M_{L}$ (the space $z$-component of the orbital angular momentum and $P_{e}$  (the permutation symmetry which can be either symmetric or antisymmetric under different sets of spin variables). Spin-isospin wavefunctions as denoted by $V_{M_{T},M_{S},\kappa}(P_{t},T,S)$ correspond to definite values of spin-$S$, isospin-$T$ and their $z$-th components ($M_{S}$ and $M_{T}$), respectively. This function transform according to the $P_{t}$ (the permutation symmetry which can be either symmetric or antisymmetric under different sets of isospin variables) representation of $S_3$ group and $\kappa$ is the row-number of $P_{t}$ \cite{Derrick1958}. 

Explicit expressions for these total angular momentum-isospin wave functions as used in Eq. (\ref{wave}) are given by \cite{thesis}:

\begin{eqnarray}
\mathcal{Y}_{1} & = & Y^0_0(s,0) V_{-\frac{1}{2} \frac{1}{2} 1} \left(a,\frac{1}{2},\frac{1}{2}\right), \nonumber \\
\mathcal{Y}_{8,1} & = & 10^{-\frac{1}{2}}\left[2 Y^2_2(s,0) V_{-\frac{1}{2} -\frac{3}{2} 1} \left(m,\frac{1}{2},\frac{3}{2}\right) \right.- \left. 3^{\frac{1}{2}}Y^1_2(s,0) V_{-\frac{1}{2} -\frac{1}{2} 1} \left(m,\frac{1}{2},\frac{3}{2}\right) \right. \nonumber \\ 
&+& \left. 2^{\frac{1}{2}}Y^0_2(s,0) V_{-\frac{1}{2} \frac{1}{2} 1} \left(m,\frac{1}{2},\frac{3}{2}\right)\right.- \left. Y^{-1}_2(s,0) V_{-\frac{1}{2} \frac{3}{2} 1} \left(m,\frac{1}{2},\frac{3}{2}\right) \vphantom{12} \right], \nonumber \\
\mathcal{Y}_{8,2} & = & 10^{-\frac{1}{2}}\left[2 Y^2_2(s,0) V_{-\frac{1}{2} -\frac{3}{2} 2} \left(m,\frac{1}{2},\frac{3}{2}\right)\right.- \left. 3^{\frac{1}{2}}Y^1_2(s,0) V_{-\frac{1}{2} -\frac{1}{2} 2} \left(m,\frac{1}{2},\frac{3}{2}\right) \right. \nonumber \\ 
&+& \left. 2^{\frac{1}{2}}Y^0_2(s,0) V_{-\frac{1}{2} \frac{1}{2} 2} \left(m,\frac{1}{2},\frac{3}{2}\right)\right. -\left.Y^{-1}_2(s,0) V_{-\frac{1}{2} \frac{3}{2} 2} \left(m,\frac{1}{2},\frac{3}{2}\right)\right],\nonumber\\
\mathcal{Y}_{9,1} & = & 10^{-\frac{1}{2}}\left[2 Y^2_2(s,2) V_{-\frac{1}{2} -\frac{3}{2} 1} \left(m,\frac{1}{2},\frac{3}{2}\right) \right.- \left. 3^{\frac{1}{2}}Y^1_2(s,2) V_{-\frac{1}{2} -\frac{1}{2} 1} \left(m,\frac{1}{2},\frac{3}{2}\right) \right. \nonumber \\ 
&+& \left. 2^{\frac{1}{2}}Y^0_2(s,2) V_{-\frac{1}{2} \frac{1}{2} 1} \left(m,\frac{1}{2},\frac{3}{2}\right)\right.- \left.Y^{-1}_2(s,2) V_{-\frac{1}{2} \frac{3}{2} 1} \left(m,\frac{1}{2},\frac{3}{2}\right) \right], \nonumber\\
\mathcal{Y}_{9,2} & = & 10^{-\frac{1}{2}}\left[2 Y^2_2(s,2) V_{-\frac{1}{2} -\frac{3}{2} 2} \left(m,\frac{1}{2},\frac{3}{2}\right)\right. - \left. 3^{\frac{1}{2}}Y^1_2(s,2) V_{-\frac{1}{2} -\frac{1}{2} 2} \left(m,\frac{1}{2},\frac{3}{2}\right) \right. \nonumber \\ 
&+& \left. 2^{\frac{1}{2}}Y^0_2(s,2) V_{-\frac{1}{2} \frac{1}{2} 2} \left(m,\frac{1}{2},\frac{3}{2}\right)\right.- \left. Y^{-1}_2(s,2) V_{-\frac{1}{2} \frac{3}{2} 2} \left(m,\frac{1}{2},\frac{3}{2}\right) \right],\nonumber\\
\mathcal{Y}_{10,1} & = & 10^{-\frac{1}{2}}\left[2 Y^2_2(a,2) V_{-\frac{1}{2} -\frac{3}{2} 2} \left(m,\frac{1}{2},\frac{3}{2}\right)\right. - \left. 3^{\frac{1}{2}}Y^1_2(a,2) V_{-\frac{1}{2} -\frac{1}{2} 2} \left(m,\frac{1}{2},\frac{3}{2}\right) \right. \nonumber \\ 
&+& \left.( 2^{\frac{1}{2}}Y^0_2(a,2) V_{-\frac{1}{2} \frac{1}{2} 2} \left(m,\frac{1}{2},\frac{3}{2}\right)\right. -\left. Y^{-1}_2(a,2) V_{-\frac{1}{2} \frac{3}{2} 2} \left(m,\frac{1}{2},\frac{3}{2}\right) \right] ,\nonumber \\
\mathcal{Y}_{10,2} & = & -10^{-\frac{1}{2}}\left[2 Y^2_2(a,2) V_{-\frac{1}{2} -\frac{3}{2} 1} \left(m,\frac{1}{2},\frac{3}{2}\right)\right.- \left. 3^{\frac{1}{2}}Y^1_2(a,2) V_{-\frac{1}{2} -\frac{1}{2} 1} \left(m,\frac{1}{2},\frac{3}{2}\right) \right. \nonumber \\ 
&+& \left.( 2^{\frac{1}{2}}Y^0_2(a,2) V_{-\frac{1}{2} \frac{1}{2} 1} \left(m,\frac{1}{2},\frac{3}{2}\right)\right. - \left.Y^{-1}_2(a,2) V_{-\frac{1}{2} \frac{3}{2} 1} \left(m,\frac{1}{2},\frac{3}{2}\right) \right]
\end{eqnarray}
and
\begin{eqnarray}
V_{-\frac{1}{2} \frac{3}{2} 1}\left(m,\frac{1}{2},\frac{3}{2}\right) = q_6 p_4,~
V_{-\frac{1}{2} \frac{3}{2} 2}\left(m,\frac{1}{2},\frac{3}{2}\right) = q_6 p_5,\nonumber\\ 
V_{-\frac{1}{2} \frac{1}{2} 1}\left(m,\frac{1}{2},\frac{3}{2}\right) = q_3 p_4,~
V_{-\frac{1}{2} \frac{1}{2} 2}\left(m,\frac{1}{2},\frac{3}{2}\right) = q_3 p_5, \nonumber\\
V_{-\frac{1}{2} -\frac{1}{2} 1}\left(m,\frac{1}{2},\frac{3}{2}\right) = q_7 p_4,~
V_{-\frac{1}{2} -\frac{1}{2} 2}\left(m,\frac{1}{2},\frac{3}{2}\right) = q_7 p_5, \nonumber \\
V_{-\frac{1}{2} -\frac{3}{2} 1}\left(m,\frac{1}{2},\frac{3}{2}\right) = q_8 p_4,~ 
V_{-\frac{1}{2} -\frac{3}{2} 2}\left(m,\frac{1}{2},\frac{3}{2}\right) = q_8 p_5,\nonumber
\end{eqnarray}
\noindent where $q_1-q_8$ are the eight eigenstates of the $S^2$ and $S_{z}$ for a system of three indistinguishable particles as $S=\frac{1}{2}$ or $s=\frac{3}{2}$ and these are determined using the double Clebsch-Gordan series. $p_1-p_8$ are the eigenstates of  $T^2$ and $T_{z}$ having the same permutation properties as the corresponding $q's$ \cite{McMillan1967}. 

The admixture of various states in the ground state of triton is attributed to the two-particle tensor operator ${\bf{S_{ij}}}$ which is defined as below:
\begin{eqnarray}
{\bf{S_{ij}}}=\frac{3\left({\boldsymbol{\sigma}}_{i}.\bf{r_{ij}}\right)\left(\boldsymbol{\sigma}_{j}.\bf{r_{ij}}\right)}{r_{ij}^2}-\boldsymbol{\sigma}_i.\boldsymbol{\sigma}_j
\end{eqnarray}
So, considering the following form of the two-nucleon potential $V_{12}(r_{ij})$ involving  the singlet spin potential  $V^{S}(r_{ij})$, the central triplet potential  $V^{ct}(r_{ij})$ and the tensor potential  $V^{t}(r_{ij})$ 
\begin{eqnarray}
V_{ij}(r_{ij})&=&\frac{V^{S}(r_{ij})}{4}(1-\boldsymbol{\sigma}_{i}.\boldsymbol{\sigma}_{j})+\frac{V^{ct}(r_{ij})}{4}(3+\boldsymbol{\sigma}_{i}.\boldsymbol{\sigma}_{j}) +{\bf{S_{ij}}}V^{t}(r_{ij})
\end{eqnarray}
while assuming $f_1(r)$ and $f_{13}(r)$ as a symmetric functions of single variable \cite{Feshbach1955} $$r=\frac{1}{2}\sum_{i,j}r_{ij}~ \mathrm{with}~ i\neq j ~\mathrm{and}~ i,j=1,2,3.$$

Employing the variational principle to extremize action
\begin{eqnarray}
\delta \int L d \tau=0,
\end{eqnarray}
the coupled differential equations for the symmetric $S$- and $D$- state components in the ground state of triton are obtained as \cite{McMillan1967}:
\begin{eqnarray}
\label{coupled}
&-&\frac{\hbar^2}{m^{\prime}}\left(\frac{d^2u}{dr^2} - \frac{15u}{4r^2}\right)+\mathcal{V^+}(r)u+\tilde{\mathcal{V}}^t(r)w =Eu \nonumber\\
&-&\frac{\hbar^2}{m^{\prime}}\left(\frac{d^2w}{dr^2} - \frac{63w}{4r^2}\right)+(\mathcal{V}^{ct}(r)-\mathcal{V}^t(r))w  +\tilde{\mathcal{V}}^t(r)u =E w 
\end{eqnarray}
where,
\[ \mathcal{V}^{+}(r) =24  \int_0^1  \left(\frac{z^4}{6}-z^3+ z^2\right)\left(\frac{V^{s}(rz)+V^{ct}(rz)}{2}\right)dz, \]
\begin{eqnarray}
 \mathcal{V}^{ct}(r) = \frac{2016}{55} \int_0^1  \left(\frac{3 z^8}{28}-\frac{7 z^7}{6}+4 z^6-\frac{20 z^5}{3}+\frac{35 z^4}{6}-3z^3 \right.+\left.z^2 \right) V^{ct}(rz) dz,\nonumber
 \end{eqnarray}
 \begin{small}
\[ \mathcal{V}^t(r) = \frac{672}{5}  \int_0^1   \left(\frac{9 z^8}{154}-\frac{212 z^7}{385}+\frac{597 z^6}{385}-2 z^5+z^4\right) V^{t}(rz) dz,\]
\end{small}
\begin{small}
\begin{eqnarray} 
\tilde{\mathcal{V}}^t(r)= \frac{336}{5} \sqrt{\frac{21}{55}} \int_0^1   \left(\frac{z^6}{6}-z^5+z^4\right) V^{t}(rz) dz
\label{term}
\end{eqnarray}
\end{small}
\noindent and $m^{\prime}=\frac{14}{15}m$ with $m$ being the mass of the nucleon and, $u(r)$ and $w(r)$ can be recognized as the following form:
 \begin{eqnarray}
 f_{1}(r)=\sqrt{\frac{30}{7}}\frac{u(r)}{r^{5/2}}, \qquad 
 f_{13}(r)=\sqrt{\frac{81}{22}}\frac{w(r)}{r^{9/2}}. 
 \end{eqnarray}
The normalization condition has been reduced to:
\begin{eqnarray}
\int_{0}^{\infty} \left[u^2(r)+w^2(r)\right]dr=1. \nonumber
\end{eqnarray}
Although $^{3}S_1$ and $^{3}D_1$ states in deuteron and  $^{2}S_{1/2}$ and $^{4}D_{1/2}$ states in triton are coupled by the tensor forces, yet one can compare  equations (\ref{coupled}) with the coupled equations for the deuteron's $^{3}S_1$ and $^{3}D_1$ components. In the case of triton, the centrifugal terms comprises of  $(\frac{3}{2}+l)(\frac{3}{2}+l+1)$ for $l=0,2$ rather than $l(l+1)$. Also, the potential energy terms in the triton's $S$-state equation includes the singlet spin potential which does not appear in the case of deuteron.

\section{Semiclassical Hamiltonian}

\noindent
Defining $\Psi= \left[
\begin{array}{c}
 u \\
 w
\end{array}
\right]$ and 
$
\textbf{V}(r)=
\left[
\begin{array}{cc}
 \frac{15 \hbar ^2}{4 m^{\prime} r^2}+\frac{15 \mathcal{V}^+(r)}{14} & \frac{15 \tilde{\mathcal{V}}^t(r)}{14} \\
 \frac{15 \tilde{\mathcal{V}}^t(r)}{14} & \frac{63 \hbar ^2}{4 m^{\prime} r^2}+\frac{15 (\mathcal{V}^{ct}(r)-\mathcal{V}^t(r))}{14}
\end{array}
\right]$

\noindent we can express Eqs (\ref{coupled}) in a matrix form 
\begin{eqnarray}
\textbf{H}\Psi=\textbf{E}\Psi,
 ~\mathrm{with}~
 \textbf{H} = \left(-\frac{\hbar ^2}{m^{\prime}}\frac{d^2}{dr^2}\right)\textbf{I} + \textbf{V}(r).
\end{eqnarray}
Since \textbf{V}(r) is a real-symmetric matrix,  it can be diagonalized  using a  general orthogonal matrix $\textbf{O}(r)$:
\begin{eqnarray}
\textbf{O}(r)=
\left[
\begin{array}{cc}
 \cos (\theta(r) /2) & \sin (\theta(r) /2) \\
 -\sin (\theta(r) /2) & \cos(\theta(r) /2) \\
\end{array}
\right],~\mathrm{where}~
\nonumber
\end{eqnarray}
\begin{eqnarray}
\tan \theta(r) = \frac{ \tilde{\mathcal{V}}^t(r)}{\frac{56\hbar ^2}{10 m^{\prime} r^2}+\frac{1}{2}(\mathcal{V}^{ct}(r)-\mathcal{V}^+(r) -\mathcal{V}^t(r)) }. \nonumber
\end{eqnarray}
\indent which leads to
\begin{equation}
\textbf{O}^{T}(r) \textbf{V}(r) \textbf{O}(r) = \left[
\begin{array}{cc}
 v_{+}(r) & 0 \\
 0 & v_{-}(r) 
\end{array}
\right] = \textbf{v}(r).
\end{equation}

\noindent Potential energy surfaces $\{v_{+}(r), v_{-}(r)\}$ are found to be of binding and scattering type respectively, when plotted in the Fig. 1. These surfaces do not intersect each other in real space but as the system evolves on both these surfaces, they are connected at a point $r_0$ (determined from $v_{+}(r)=v_{-}(r)$) which could be complex. This would mean that the system tunnels to the other surface from where it is scattered back. 
\begin{figure}[h]
\includegraphics[scale=0.5]{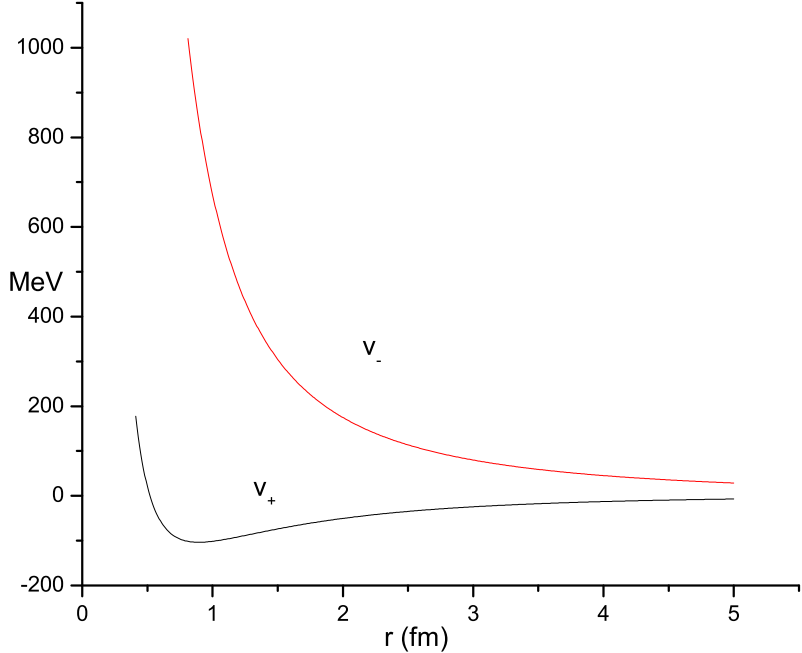}
\caption{Binding-type potential energy surface $v_{+}(r)$ and scattering-type potential energy $v_{-}(r)$ are plotted w.r.t. $r$ for Feshbach-Pease Potential No. 1. They intersect each other in the complex plane at (5.48242 + 11.4104 i)  fm. }
\end{figure}

%
%
The idea is to apply the orthogonal transformation and similarity transform $H$ to $H^{\prime}$. Now we encounter a term like ${\textbf{O}^T(r)}p_r^2{\textbf{O}(r)}=({\textbf{O}^T(r)}p_r^2{\textbf{O}(r)})^2$. Due to non-commutability of $\textbf{O}(r)$ and $p_{r}$, the term becomes $(p_{r}-\hbar A(r))^2$ with an appropriately identified gauge potential $A(r)$. Thus, we get terms of $\mathcal{O}(1)$, $\mathcal{O}(\hbar)$, $\mathcal{O}(\hbar^2)$ where $\mathcal{O}(1)$ terms are diagonal. To the non-diagonal part, we can again apply repeat the above procedure and arrive at diagonal term such that the next correction is at $\mathcal{O}(\hbar^2)$ \cite{gaspard}. This gives us two semiclassical wave equations, 
\begin{equation}
h^{\pm}\psi^{\pm}=E\psi^{\pm} ~ \mathrm{with} \nonumber
\label{eigen}
\nonumber
\end{equation}
\begin{eqnarray}
 ~ h^{\pm} & = &-\frac{\hbar^2}{m^{\prime}}\left[1\pm \frac{\hbar^2}{m^{\prime}(v_+(r) - v_-(r))}\left(\frac{d\theta(r)}{dr}\right)^2\right] \frac{d^2}{dr^2}+v_\pm(r)+\frac{\hbar^2}{8m^{\prime}}
\left[\frac{d\theta(r)}{dr}\right]^2, \nonumber
\label{hamiltonian}
\end{eqnarray}
are the two semiclassical Hamiltonians  yielding eigenvalue $E$  when operated upon by the two  combinations of $u(r)$ and $w(r)$ defined as follows:

\begin{eqnarray}
\psi^{+}(r)&=& u(r) \cos \frac{\theta (r)}{2}-w(r) \sin \frac{\theta (r)}{2} ~\mathrm{and}\nonumber\\
\psi^{-}(r)&=& w(r) \cos \frac{\theta (r)}{2}+u(r) \sin \frac{\theta (r)}{2}. 
\label{solution}
\end{eqnarray}
One may notice in  (\ref{hamiltonian}), the ground state of triton involves two ``mass states" with the following  effective masses:
\begin{eqnarray}
M_{\pm}(r)=m^{\prime}\left[1\pm\frac{\hbar^2}{m^{\prime}(v_+(r) - v_-(r))}\left(\frac{d\theta(r)}{dr}\right)^2\right]^{-1}.\nonumber
\end{eqnarray}
These states are separated by an energy difference of about 2.5 MeV in triton when $r$=1 fm.
\begin{figure}[h]
\includegraphics[scale=0.35]{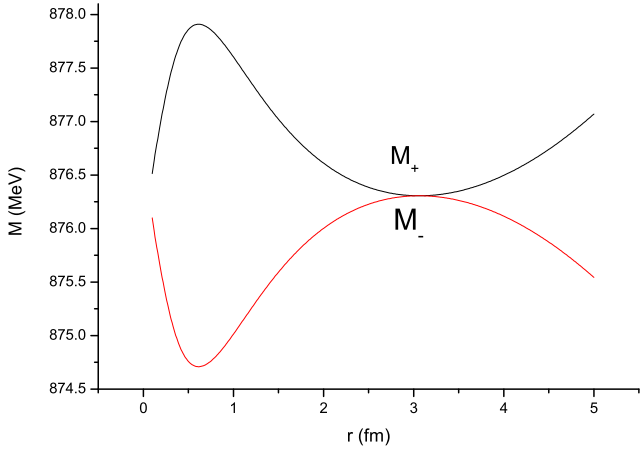}
\caption{$M_{\pm}(r)$ are plotted w.r.t. $r$ in this figure. Mass states are separated by an energy difference of about 2.5 MeV in triton when $r$=1 fm.}
\end{figure}

\section{Results and Discussion}

We have used the following forms of Yukawa type potentials to determine the expressions (\ref{term}):
\begin{small}
\begin{eqnarray}
V^{s}(r)= V_{os} \frac{e^{-\mu_{s} r}}{\mu_{s} r}, 
V^{ct}(r)= V_{ot} \frac{e^{-\mu_{t} r}}{\mu_{t} r}, 
V^{t}(r)= V_{ote} \frac{e^{-\mu_{te} r}}{\mu_{te} r}.
\label{potential}
\end{eqnarray}
\end{small}
On the basis of the values chosen for the parameters $(V_{os}, V_{ot}, V_{ote},  \mu_{s}, \mu_{t}, \mu_{te} )$ as listed in the Table I, potentials in  (\ref{potential}) can be categorized into four types of Feshbach-Pease potentials \cite{pease}.

\begin{table}[h]
\centering
\begin{tabular}{c c c c c c c}
\hline
Feshbach- & & & & & &\\
Pease & $-V_{os}$  & $-V_{ot}$  & $-V_{ote}$  &  $\mu_{s}$  &  $\mu_{t}$ & $\mu_{te}$ \\ 
 Potential& (MeV)  & (MeV) & (MeV) &  (fm$^{-1}$)  &  (fm$^{-1}$) & (fm$^{-1}$)  \\
\hline
 No. 1   & 55.0323                                            & 65.1839                                            & 26.3907                                             & 0.84459 & 0.84459 & 0.4713 \\ \hline
 No. 2   & 55.0323                                            & 59.3456                                            & 34.7892                                             & 0.84459 & 0.84459 & 0.5474 \\ \hline
 g=0    & 55.0323                                            & 55.0323                                            & 43.2731                                             & 0.84459 & 0.84459 & 0.5907 \\ \hline
 No. 3   & 55.0323                                            & 48.4469                                            & 49.5169                                             & 0.84459 & 0.84459 & 0.6523 \\
 \hline
 \label{TableYukawa}
\end{tabular}
\caption{Parameters for different Feshbach-Pease Potentials used for the calculations \cite{pease}.}
\end{table}

\subsection{Ground state of triton}

Now we will perform the realistic calculations to determine the form of symmetric components of spatial wavefunction for the ground state of triton which occurs at $E=-8.482$ MeV \cite{audi}. Semiclassical wave equation (\ref{eigen}) can be written in the following simplified manner upto $\mathcal{O}(\hbar^2)$ as:
\begin{eqnarray}
\frac{d^2 \psi^{+}(r)}{dr^2}+(k^2-U(r))\psi^{+}(r)=0 
\label{diff}
\end{eqnarray} 
\noindent where $$k=\frac{m^{\prime}E}{\hbar^2}, \qquad \mathrm{and} ~ U(r)=\frac{m^{\prime}}{\hbar^2}v_{+}(r)+\frac{1}{2}\left[\frac{d\theta}{2dr}\right]^2.$$
Expanding $U(r)$ in Taylor series about turning point $r_2$ and retaining only leading order
\begin{eqnarray}
\frac{d^2 \psi^{+}(r)}{dr^2}=(r-r_2)U^{\prime}(r)\psi^{+}(r). 
\label{Airy}
\end{eqnarray}
as $U(r_2)=k^2$. This the well-known Airy's differential equation whose analytic solution is  given below \cite{brack2003}: 
\begin{equation}
\psi ^{+}(r) = \mathrm{Ai}[U'(r_2)^{1/3} (r-r_2)].
\end{equation}
We rewrite $h^-\psi^-=E\psi^-$ too in the form of the second-order differential equation with
\begin{eqnarray}
U_1(r)= \frac{m^{\prime}}{\hbar^2}v_-(r)+\frac{1}{2}\left(\frac{d\theta (r)}{2dr}\right)^2
\end{eqnarray}
and evaluate $\psi^{-}(r)$ exploiting the method as described in \cite{bender}:
\begin{equation}
\psi ^{-}(r) = \mathrm{Re}[\mathrm{Ai}(U'(r_3)^{1/3} (r-r_3))].
\end{equation}
where $r_3=(-1.1078+2.1338 i)~\mathrm{fm}$ is the complex turning point found using Feshbach-Pease Potential No. 1. Of course, there are other complex turning points too but in the present context only  turning point $r_3$ has been used.

The radial wavefunctions, $u(r)$ and $w(r)$ can be readily evaluated using the following relations:
\begin{eqnarray}
u(r)& = & \psi^{+}(r) \cos(\theta (r) /2) + \psi_{-}(r) \sin (\theta (r) /2) \nonumber\\
w(r) &= & \psi^{-}(r) \cos(\theta (r) /2) - \psi_{+}(r) \sin (\theta (r) /2)
\end{eqnarray}
 
\begin{figure}[h]
\includegraphics[scale=0.35]{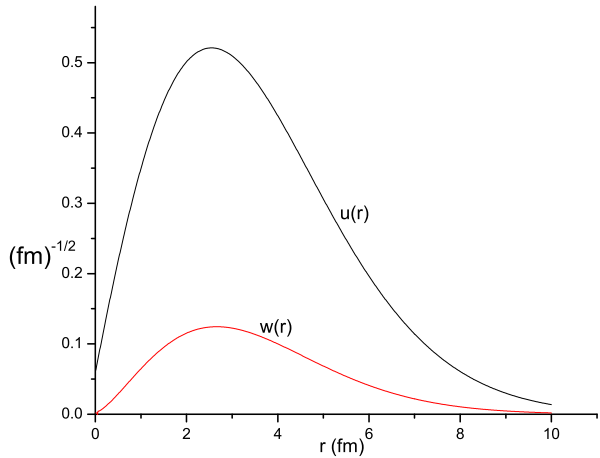}
\caption{Radial components of the wavefunctions $(u(r)~\mathrm{w(r)})$ for $S$- and $D$- states of triton are plotted using Feshbach-Pease Potential No. 1. An almost similarity with the numerical results \cite{McMillan1967} is observed.}
\end{figure}

\noindent and the obtained results for the radial component of the wavefunction are plotted in Fig. 3. Similarly, for other types of  Feshbach-Pease Potentials, we find $u(r)$ and $w(r)$ and almost similar trends are obtained. 

Consequently, we  determine the $D$- state probability $(P_D)$ as
\begin{equation}
P_D = \int_0^\infty w^2(r) dr.
\end{equation} 
and the results obtained show better agreement with the experimental value as compared to the numerically  obtained ones with earlier methods (Table II). Best-fitting exponential trial wavefunction for Feshbach-Pease potentials has been used to perform the calculations under ``FP''-category. However,  to simplify the coupled second-order differential equations (\ref{coupled}), by ``Modified Feshbach-Rubinow (FR)"  the radial components of the wavefunctions are related   as following:
\begin{eqnarray}
u(r)=\alpha w(r),
\label{relation}
\end{eqnarray}
with $\alpha$ is a constant. While, Runge-Kutta method has been used to solve the coupled differential equations following the conditions as described in \cite{thesis} (coupled Feshbach-Rubinow (FR) Method).

\begin{table}[h]
\centering
\begin{tabular}{c c c c c c }
\hline
Feshbach-Pease & & & $P_{D}$(\%)   & & \\
Potential Type & & & &  \\
\hline\\
&  FP  &  Coupled  & Modified  & Semiclasssical &  \\
& & FR  & FR & method & \\
\hline
 No. 1   & 2.2 & 1.9   & 1.6 & 4.9 \\ \hline
 No. 2   & 2.8 & 2.8   & 1.9 & 4.5 \\ \hline
  g=0   & 3.1 & 3.1   & 2.4 & 4.4 \\ \hline
 No. 3   & 3.6 & 3.6   & 2.4 & 4.0 \\ \hline
\end{tabular}
\caption{D-State probability $(P_{D}(\%))$ obtained for four types of Feshbach-Pease Potentials. Exponential trial wavefunction has been used under FP-method. By Coupled FR-method, calculations are done using Runge-Kutta method with conditions as described in \cite{thesis}. Modified FR-method involves the simplification using (\ref{relation}). Our semiclassical results are in agreement with the values  concluded from the magnetic moment calculations i.e. 4-5\% \cite{blatt}.}
\end{table}

\section{Concluding remarks}

Exact understanding, even if semiclassical, of the three-body problem has been a longstanding challenge. Initial efforts towards the explanation of the ground state of triton has been made by Derrick et al. \cite{Derrick1958, derrick1960a, derrick1960b}. Also, triton's wavefunction and energy has been obtained using non-variational approach \cite{McMillan1965, Feshbach1955}. A large literature based on adaptation of Faddeev equations and ab initio methods lead us to useful numerical results. Our approach has been essentially analytical, and has been systematic as we obtain the final result by organizing the analysis of Hamiltonian operator and the corresponding wave equation at different orders of the Planck constant. We have determined the symmetric components of the spatial part of $S$- and $D$- states' wavefunctions for triton $(^{3}H)$. We have solved the coupled differential equations for the two admixed states $^{2}S_{1/2}$ and $^{4}D_{1/2}$ employing semiclassical methods. It has been shown by the action quantization that the ground state actually exists at $E=-8.482$ MeV in triton. The probability for the existence of the of $D$-state is found to be almost 4\% to 5\%. Such an agreement with the experimentally deduced values testifies the correctness of the form of the wavefunctions found by us. With this solution, we realize that the semiclassical matrix method can be used to find states of few-body systems interacting via tensor forces. Indeed, there have been a number of attempts to obtain a solution for $^4$He (see, for instance, \cite{adhikari}).


\section*{References}
\begin{thebibliography}{30}
\bibitem{blatt} Blatt J and Weisskopf V 1979 {\it{Theoretical Nuclear Physics}} (Springer New York).
\bibitem{Jain2004} Jain S R 2004 {\it{J. Phys. G: Nucl. Part. Phys.}} {\bf{30}} 157.
\bibitem{Derrick1958} Derrick G and Blatt J 1958 {\it{Nucl. Phys.}} {\bf{8}} 310.
\bibitem{mitra} Mitra A N 1962 {\it{Nucl. Phys.}} {\bf{32}} 529.
\bibitem{gibson67} Gibson B F 1967 {\it{Nucl. Phys. B }} {\bf{2}} 501.
\bibitem{gibson85}  Chen C R, Payne G L, Friar J L and Gibson B F 1985 {\it{Phys. Rev. C}} {\bf{31}} 2266.
\bibitem{wu} Wu Y, Ishikawa S and  Sasakawa T 1993 {\it{Few-body systems}} {\bf{15}} 145.
\bibitem{berry} Berry M V 1978 {\it{ed. S Jorna, Am. Inst. Ph. Conf. Proc}} {\bf{46}} 16.
\bibitem{gutzwiller} Gutzwiller M C 1998 {\it{Rev. Mod. Phys.}} {\bf{70}} 589.
\bibitem{littlejohn} Littlejohn R G and Weigert S 1993 {\it{Phys. Rev. A}} {\bf{48}} 924.
\bibitem{gaspard} Gaspard P, Alonso D and Burghardt I 1997 {\it{Adv. Chem. Phys.}} {\bf{90}} 105.
\bibitem{brack2003} Brack M and Bhaduri R 2003 {\it{Semiclassical Physics, Frontiers in Physics}} (Westview Press).
\bibitem{Jain1998}  Jain S R and Pati A K 1998 {\it{Phys. Rev. Lett.}} {\bf{80}} 650.
\bibitem{Jain1995} Brack M and Jain S R 1995 {\it{Phys. Rev.A}} {\bf{51}} 3462.
\bibitem{Kaur} Kaur H and Jain S R 2015 {\it{J. Phys. G: Nucl. Part. Phys.}} {\bf{42}} 115103.
\bibitem{thesis}Best M E 1966 {\it{On the symmetric S- and D- components of the triton wavefunction}} (The University of British Columbia).
\bibitem{McMillan1967} McMillan M and Best M 1967 {\it{Nuclear Physics A}} {\bf{105}} 649.
\bibitem{Feshbach1955} Feshbach H and Rubinow S I 1955 {\it{Phys. Rev.}} {\bf{98}} 188.
\bibitem{pease}  Pease R L and Feshbach H 1952 {\it{Phys. Rev.}} {\bf{88}} 945
\bibitem{audi}  Audi G, Kondev F G, Wang M, Pfeiffer B, Sun X, Blachot J and  MacCormick M 2012 {\it{Chinese Physics C}} {\bf{36}} 1157.
\bibitem{bender} Bender C M and  Jones H F 2012 {\it{Phys. Rev. A}} {\bf{85}} 052118.
\bibitem{derrick1960a} Derrick G 1960 {\it{Nuclear Physics}} {\bf{16}} 405.
\bibitem{derrick1960b} Derrick G  1960 {\it{Nuclear Physics}} {\bf{18}} 303.
\bibitem{McMillan1965} McMillan M 1965 {\it{Canadian Journal of Physics}} {\bf{43}} 463.
\bibitem{adhikari} Adhikari S K, Frederico T, Goldman I D and Sharma S S 1994 {\it{Phys. Rev. C}} {\bf{50}} 822.
\end{thebibliography}
\end{document}